\documentclass{article}

\usepackage{arxiv}

\usepackage[utf8]{inputenc} 
\usepackage[T1]{fontenc}    
\usepackage{hyperref}       
\usepackage{url}            
\usepackage{booktabs}       
\usepackage{amsfonts}       
\usepackage{nicefrac}       
\usepackage{microtype}      
\usepackage{lipsum}
\usepackage{graphicx}
\usepackage{xcolor}
\usepackage{multirow}
\usepackage{tcolorbox}

\graphicspath{ {./images/} }

\title{Statistical hypothesis testing versus machine-learning binary classification: distinctions and guidelines}

\author{
 Jingyi Jessica Li \\
  Department of Statistics\\
  University of California\\
  Los Angeles, CA 90095-1554 \\
  \texttt{jli@stat.ucla.edu} \\
   \And
 Xin Tong \\
  Department of Data Sciences and Operations\\
  Marshall School of Business\\
  University of Southern California\\
  Los Angeles, CA 90089 \\
  \texttt{xint@marshall.usc.edu} \\
}

\begin{document}
\maketitle
\begin{abstract}
Making binary decisions is a common data analytical task in scientific research and industrial applications. In data sciences, there are two related but distinct strategies: hypothesis testing and binary classification. In practice, how to choose between these two strategies can be unclear and rather confusing. Here we summarize key distinctions between these two strategies in three aspects and list five practical guidelines for data analysts to choose the appropriate strategy for specific analysis needs. We demonstrate the use of those guidelines in a cancer driver gene prediction example.   
\end{abstract}


\section{Introduction}
Making binary decisions is one of the most common human cognitive activities. Binary decisions are everywhere: from spam detection in IT technologies to biomarker identification in medical research. For example, facing the current COVID-19 pandemic, medical doctors need to make a critical binary decision: whether an infected patient needs hospitalization. Living in a big data era, how can we make rational binary decisions from massive data? 

In data sciences, two powerful strategies have been developed to assist binary decisions: the statistical \textit{hypothesis testing} \cite{lehmann2006testing} and the machine-learning \textit{binary classification} \cite{murphy2012machine}. While both strategies are popular and have achieved profound successes in various applications, their distinctions are largely obscure to practitioners and even data scientists sometimes. An important reason is that the two strategies are usually introduced in different classes and covered by different textbooks, with few exceptions such as \cite{wasserman2013all}. Another source of confusion is the ambiguous use of the term ``test'' to represent both strategies in our daily life, such as in ``statistical test'' and ``COVID-19 test,'' where the latter is, in fact, binary classification and will be referred to as ``COVID-19 diagnosis" in this work.

There are online discussions about the relationship between binary classification and hypothesis testing; however, they focus on specific cases and are not unified into a coherent picture. For example, one discussion compares the Student's $t$ test, a specific statistical test, with the support vector machines, a specific binary classification algorithm \cite{MaxG:2016}. Another discussion compares the asymmetric nature of hypothesis testing with the general lack of asymmetry in binary classification \cite{icurays1:2016}. Besides online discussion, there are research works that borrow ideas from hypothesis testing to develop binary classification algorithms \cite{Liao.Akritas.2007, He.Sheng.Liu.Zou.2019}, but these works do not aim to link or compare the two strategies. 

 Here we attempt to make the first efforts to summarize key distinctions between hypothesis testing in frequentist statistics \footnote{In this work, we only discuss hypothesis testing under the frequentist framework in statistics, and our disucssion does not pertain to Bayesian hypothesis testing \cite{efron2016computer}.} and binary classification in machine learning. We also provide five practical guidelines for data analysts to choose between the two strategies. In our discussion, we will frequently use biomarker detection and disease diagnosis as examples of hypothesis testing and binary classification, respectively. In these two examples, \textit{instances}\footnote{Instances are often referred to as ``individuals'' in biomedical sciences, ``objects'' in engineering, ``observations'' in statistics, and ``data points'' in data sciences. Although many researchers outside of statistics refer to instances as ``samples'' (in the plural form), here we stick with the classic statistical definition: a ``sample'' is a collection of instances.} refer to patients, and \textit{features}\footnote{Features are also referred to as ``variables'' and ``covariates'' in statistics.} refer to patients' diagnostic measurements such as blood pressure and gene expression levels.

\section{Distinctions between hypothesis testing and binary classification}
\label{sec:distinction}

Hypothesis testing and binary classification are rooted in two different cultures: \textit{inference} and \textit{prediction}, each of which has been extensively studied in statistics and machine learning respectively in the historical development of data sciences \cite{breiman2001statistical}. Briefly, an inferential task aims to infer an unknown truth from observed data, and hypothesis testing is a specific framework whose inferential target is a binary truth, i.e., an answer to a yes/no question. For example, deciding whether a gene is an effective COVID-19 biomarker in the blood is an inferential question, whose answer is unobservable.
In contrast, a prediction task aims to predict an unobserved property of an instance, such as a patient or an object, based on the available features of this instance. Such prediction relies on building a trustworthy relationship, i.e., a \textit{prediction rule}, from the input features to the target property, which must be based on human knowledge (throughout the human history) and/or established from data (after computing devices were developed). Binary classification is a special type of prediction whose target property is binary, and COVID-19 diagnosis is an example. In screening patients for COVID-19 exams, medical doctors make binary decisions based on patients' symptoms (input features), and their decision rules are learned from previous patients' diagnostic data and medical literature.

Hypothesis testing is built upon the concept of \textit{statistical significance}, which intuitively means that the data we observe present strong evidence against a presumed \textit{null hypothesis}, the default. In the example of testing whether a gene is a COVID-19 biomarker in blood, the null hypothesis is that this gene does not exhibit differential expression in the blood of uninfected individuals and COVID-19 patients. This formulation reflects a conservative attitude: we do not want to call the gene a biomarker unless its expression difference is large enough between the healthy and diseased patients we measured. Statistical hypothesis testing provides a formal framework for deciding a threshold on the expression difference so that the gene can be identified as a biomarker with the desired confidence. A crucial fact about hypothesis testing is that the null and alternative hypotheses pertain to a property of an unseen population. As a result, we cannot know whether the null hypothesis holds or not. What we have access to is instances and their features, i.e., \textit{data}, from the population, and hypothesis testing allows us to infer how unlikely the data are generated from the null hypothesis.

In machine learning, binary classification belongs to supervised learning, as it is supervised by quality \textit{training data} that contain training instances from two classes, and each training instance is labeled as class 0 or 1 with a set of feature values. A binary decision rule is first constructed from the training data and next applied to predict unobserved binary labels of new objects from their observed feature values. Binary classification embodies a large class of algorithms that automatically learn prediction rules from training data. In an ideal scenario, a prediction rule follows a scientific law, such as in Newton's second law of motion, where the acceleration of an object is determined by the net force acting on the object and the mass of the object. However, most prediction tasks do not have scientific laws to follow, and the prediction rules learned from data could be useful but not necessarily revealing scientific truth \cite{riley2019three}. For example, we can effectively predict the coming of autumn from our observation of falling leaves, which, however, do not cause autumn to come. Nevertheless, the lack of scientific interpretation is often not a major concern in many industrial applications such as spam detection and image recognition, where prediction algorithms have achieved tremendous successes, promoting machine learning to become a spotlight discipline with broad impacts on everyone's life. Still, a necessary condition for binary classification to succeed is that training instances are good representatives of new instances to make predictions for. A notorious cautionary tale is Google Flu Trends, which mistakenly predicted a nonexistent flu epidemic because its training data did not well represent the long-term dynamics of flu outbreaks \cite{harford2014big, efron2016computer}.

We summarize the key distinctions between hypothesis testing and binary classification in three aspects: data in relation to binary decisions, construction of decision rules, and evaluation criteria. Our discussion will be centered around four concepts: \textit{binary questions}, \textit{binary answers}, \textit{decision rules}, and \textit{binary decisions}, which we define for each strategy in Table~\ref{tab:concepts}. We note that these four concepts belong to three stages in a typical data analysis: conceptual formulation (when binary questions and binary answers are formulated in a researcher's mind), analysis (when a decision rule is constructed), and conclusion (when a binary decision is made).

\begin{table}[htbp]
 \caption{Four concepts under hypothesis testing and binary classification}
  \centering
  \begin{tabular}{llp{6cm}p{6cm}}
    \toprule
    Concept     & & Hypothesis testing     & Binary classification \\
    \toprule
    Binary question  &  & Is the null hypothesis false? (unanswerable) & Does the instance have a label 1?      \\
    \midrule
    \multirow{2}{*}{Binary answer}    & 0 (No) & The null hypothesis is true (unobservable) & The instance has a label 0      \\
        & 1 (Yes) & The null hypothesis is false (unobservable) & The instance has a label 1\\
    \midrule
    Decision rule & & A statistical test that inputs data and outputs a $p$-value, which is compared against a user-specified significance level $\alpha$ & A trained classifier that inputs an unlabelled instance and outputs its predicted label\\
    \midrule
    \multirow{2}{*}{Binary decision}    & 0 & Do not reject the null hypothesis & Label the instance as 0      \\
        & 1 & Reject the null hypothesis & Label the instance as 1\\
    \bottomrule
  \end{tabular}
  \label{tab:concepts}
\end{table}

\subsection{Data in relation to binary decisions.} 
In this aspect, hypothesis testing and binary classification have two distinctions: \textbf{(1) number of instances to make one decision} given a decision rule and \textbf{(2) availability of known binary answers} in data. In hypothesis testing, each binary decision, rejecting a null hypothesis or not, is made from a collection of instances, called a \textit{sample} in statistics. For example, to investigate whether a gene is a COVID-19 biomarker in blood, a researcher needs to collect blood from multiple uninfected and infected patients, whose number is called the \textit{sample size}, and measure this gene's expression within. Then the binary decision regarding whether to call this gene an informative biomarker will be made jointly from the collected measurements. If multiple genes are tested simultaneously, we are in a situation called \textit{multiple testing} \cite{efron2012large}, which is commonly employed in large-scale exploratory studies. No matter the number of tests being one or many, the number of instances used for each test should better exceed one in the practices of hypothesis testing. In fact, the greater the number of instances, the more we trust our decisions. We will further discuss the impact of the number of instances on decision rules in the third aspect (Section~\ref{sec:third_aspect}).

In contrast, binary classification makes a binary decision for every instance that needs a binary label. In COVID-19 diagnosis, a doctor needs to decide what patients should be hospitalized, and each patient will receive one decision. In other words, the number of instances in need of binary labels equals the number of decisions. Here training instances are not counted, because they already have binary labels. In practice, binary classification can be easily confused with multiple testing, as both strategies make multiple binary decisions (see the cancer driver gene prediction example in Section~\ref{sec:example}). A way to distinguish the two strategies is to count the number of input instances used to make one decision given a decision rule, whose construction is discussed in Section~\ref{sec:construction}. 

Another distinction is the availability of known answers to binary questions in mind. Such answers are always lacking for hypothesis testing questions but available in training data for binary classification. In hypothesis testing, a binary question is regarding the validity of a null hypothesis, and the answer to this question is an unobservable truth about an unseen population. For example, we do not know a priori whether a gene is a biomarker; otherwise, we would not need to do hypothesis testing. However, the unobservable binary answer is often mistaken as a binary decision---whether or not to reject the null hypothesis---an action taken based on a sample of instances and dependent upon a decision rule (Table\ref{tab:concepts}). This mistake is commonly seen in scientific research papers that claim, ``the null hypothesis is correct (or incorrect) because the $p$-value is large (or small).'' Here we raise a strong caution against this misuse. 

Unlike in hypothesis testing, a binary question in classification is regarding the binary label of an instance, and we already have known answers (labels) for training instances, which we utilize to build a decision rule to predict labels of unlabelled instances. It is worth emphasizing that a decision rule cannot be constructed if all training instances have the same label, say 0; hence, training data must contain both binary labels. For example, doctors diagnose new patients based on previous patients' data with diagnosis decisions. In brief, hypothesis testing has no concept of training data, because data contain no answers to binary questions being asked; in contrast, training data serve as a critical component in binary classification.

\subsection{Construction of decision rules.} \label{sec:construction}
In hypothesis testing, the construction of a decision rule, also known as a \textit{statistical test}, relies on three essential components: a test statistic that summarizes the data, the distribution of the test statistic under the null hypothesis, and a user-specified significance level $\alpha$, which indicates the tolerable \textit{type I error}, i.e., the conditional probability of mistakenly rejecting the null hypothesis given that it holds. The first two components lead to a $p$-value between $0$ and $1$, with a smaller value indicating stronger evidence against the null hypothesis. Then the null hypothesis is rejected if the $p$-value does not exceed $\alpha$, a small value set by users based on their tolerance level\footnote{In many textbooks and practices, $\alpha$ is set to $0.05$ by convention. However, we want to emphasize that this convention is just an arbitrary choice and should not be taken as the ritual. For example, there is a recent proposal to change the ``default'' value of $\alpha$ to $0.005$ \cite{benjamin2018redefine}. Both $0.05$ and $0.005$ are arbitrarily set thresholds.}. Numerous statistical tests have been developed since the advent of statistics, and a few of them, such as Student's $t$ test and Wilcoxon's rank-sum test, have become standard practices in data analysis. Due to the wide popularity and meanwhile common misuses of hypothesis testing, there are recent in-depth and extensive discussions on the proper use and interpretation of $p$-values in and outside of the statistics community \cite{wasserstein2016asa, benjamin2018redefine}. 

In multiple testing, the choice of $\alpha$ value is determined by an overall objective on all tests together, and two widely-used objectives are the \textit{family-wise error rate} (FWER, the probability of wrongly rejecting at least one null hypothesis) and the \textit{false discovery rate} (FDR, the expected proportion of falsely rejected hypotheses among all rejections) \cite{efron2016computer}\footnote{In high-throughput data analyses common in genomics and proteomics, the FDR is the most popular objective, while the FWER is rarely used due to its over-conservativeness. However, the FWER is still frequently used in scientific research where a moderate (e.g., fewer than 20) number of hypothesis test are performed together.}. The \textit{Bonferroni correction} is a conservative but guaranteed approach to control the FWER \cite{bonferroni1936teoria}. The \textit{Benjamini-Hochberg procedure} is a widely-used approach to control the FDR \cite{benjamini1995controlling}, and there is a recent approach \textit{knockoffs} to control the FDR when exact $p$-values cannot be achieved \cite{barber2015controlling, candes2016panning}\footnote{The FDR is a frequentist criterion. Under the Bayesian framework, empirical Bayes criteria, including the positive false discovery rate \cite{storey2003positive}, the local false discovery rate \cite{efron2008microarrays}, and the local false sign rate \cite{stephens2017false}, have been developed to control the number of false positives in multiple testing.}. It is worth noting that the construction of a decision rule in hypothesis testing does not necessarily require access to data. For example, in the classic Student's two-sample $t$ test, under the assumption that the two samples (sets of instances) are generated from two normal distributions, the decision rule only depends on the two sample sizes and a user-specified $\alpha$ value (Example below). When researchers have collected a gene's expression data in many diseased and healthy patients and have verified that the two samples approximately follow normal distributions, they can simply apply the two-sample $t$ test, a readily-available decision rule, to their data and decide whether this gene can be called a biomarker at their desired $\alpha$ value. If the normal distributional assumption does not seem to hold, researchers may use the Wilcoxon rank-sum test that does not have this assumption but only requires all the instances to be independent. Hence, in applications of hypothesis testing, the most critical step is to choose an appropriate statistical test, i.e., decision rule, by checking the test's underlying assumptions on data distribution. Meanwhile, the construction of valid new decision rules is mostly the job of academic statisticians. 

\begin{tcolorbox}[title=Example: Decision rule of the Student's two-sample $t$ test]
Suppose that we have two samples of sizes $10$ and $12$ from two normal distributions, and we are interested in whether the two normal distributions have the same mean. Then the null hypothesis is that the two normal distributions have the same mean, and the alternative hypothesis is the opposite. The test statistic---the two-sample $t$ statistic---follows the $t$ distribution with $20$ degrees of freedom ($t_{20}$) under the null hypothesis. Then given a significance level $\alpha \in (0,1)$, we would reject the null hypothesis if the $t$ statistic has an absolute value greater than or equal to $F^{-1}_{t_{20}}(1-\alpha/2)$, i.e., the $(1-\alpha/2)$-th quantile of the $t_{20}$ distribution. For example, if $\alpha = 0.05$, $F^{-1}_{t_{20}}(1-\alpha/2) = 2.085963$. This \textbf{decision rule} is equivalent to that the $p$-value is less than or equal to $\alpha$. Note that this decision rule does not depend on the observed $t$ statistic value based on a dataset. 
\\\\
However, a \textbf{decision} depends on an observed $t$ statistic value calculated from an actual dataset. For example, if the $t$ statistics is $3$, then we would reject the null hypothesis at $\alpha = 0.05$.
\end{tcolorbox} 

In contrast to hypothesis testing, we do not usually have available decision rules to choose from in binary classification; instead, we need to construct a decision rule from training data in most applications. Image classification and speech recognition are probably two famous exceptions, where superb decision rules (classifiers) have been trained from industry-standard massive image and speech data sets that well represent almost all possible images and speeches that need labeling (decisions) in daily applications. Yet in biomedical applications such as COVID-19 diagnosis, a good decision rule is often lacking but needed to be constructed from in-house training data that represent future local patients in need of diagnosis. Despite its reliance on quality training data that contain a reasonable number of instances with accurate binary labels, binary classification is fortunate to have access to dozens of powerful algorithms that can be directly applied to training data to construct a decision rule. Famous algorithms include the logistic regression, support vector machines, random forests, gradient boosting, and the resurgent neural networks (and its buzzword version ``deep learning'') \cite{friedman2001elements, murphy2012machine}. Same as in hypothesis testing, the most critical step in applications of binary classification is the choice of an appropriate algorithm to build a decision rule from training data, while the development of new algorithms is a focus of data science researchers. 

\subsection{Evaluation criteria for decision rules.}\label{sec:third_aspect}
Realizing the many possible ways of constructing decision rules in both hypothesis testing and binary classification, users face a challenging question in data analysis: how should I compare and evaluate decision rules? In hypothesis testing, statistical tests (decision rules) designed for the same null hypothesis are compared in terms of \textit{power}: the conditional probability of correctly rejecting the null hypothesis given that it does not hold, e.g., correctly identifying an effective biomarker. Under the same significance level $\alpha$, the larger the power, the better the test. The \textit{Neyman-Pearson lemma} provides the theoretical foundation for the most powerful test; however, in many practical scenarios, the Neyman-Pearson lemma does not apply and the most powerful test is not achievable, so statisticians have put continuous efforts into developing more powerful tests, such as in the flourishing field of statistical genetics \cite{sham2014statistical}. For users, the power of a statistical test is not observable from data, which contain no information regarding the null hypothesis being true or not. Hence, the only evaluation criterion for users to choose among many statistical tests is whether their data seem to fit each test's underlying assumptions on data distribution, which can be quite tricky sometimes and require consulting from statisticians. If many tests pass this check, most users would choose a popular test. An advanced user might opt for the test that gives the smallest $p$-value, i.e., the strongest evidence against the null hypothesis. However, this option should be used with extreme caution, as it could easily become ``p-hacking'' or data dredging if without sufficient justification \cite{head2015extent}. 

In binary classification, the evaluation criteria are more transparent and easier to understand, as they all rely on some sorts of prediction accuracy of a decision rule on \textit{validation data}, which contain binary labels, represent future instances that need labeling, and most importantly, are not part of the training data. Users may wonder: what if I only have one set of data with binary labels? A straightforward answer is to randomly split the data into training and validation cohorts, use the training cohort to construct a decision rule, and apply the rule to the validation cohort to evaluate a chosen prediction accuracy\footnote{As in the Google Flu Trends example, if the training data are not representative of future instances, this splitting idea would not work.}. This answer is the core idea leading to \textit{cross-validation}, the dominant approach for evaluating binary classification rules, and more generally, prediction rules \cite{efron2016computer}. If users prefer not to split the data due to its limited number of instances, probabilistic approaches are available, and they allow users to use the whole data set to train and subsequently evaluate a decision rule. Famous examples include the \textit{Akaike's information criterion} (AIC) and the \textit{Bayesian information criterion} (BIC) \cite{efron2016computer}. However, there is no free lunch; most of these non-splitting approaches require assumptions on data distribution\footnote{If these probabilistic assumptions do not hold, there is no guarantee how the decision rule would perform on a future instance.} and do not apply to binary classification algorithms that are not probability-based, while cross-validation has no such restrictions. In terms of prediction accuracy, the most commonly-used measure is the \textit{overall accuracy}: the percentage of correctly labeled instances in the validation data, e.g., the percentage of correctly diagnosed patients in a cohort not used for training the decision rule. In many applications where the two classes corresponding to binary labels $0$ and $1$ have equal importance, this measure is reasonable. In disease diagnosis, however, the two directions of misdiagnosis: predicting a diseased patient as healthy vs. predicting a healthy individual as diseased, are likely to have unequal importance, which would depend on the severity of the disease, the abundance of medical resources, and many other factors. For example, in countries with well-established health-care systems, diagnosis for high-mortality cancer patients should focus on reducing the \textit{false negative rate}, i.e., the chance of missing a patient with a malignant tumor; hence, a more relevant prediction accuracy would be the \textit{true positive rate} (one minus the false negative rate) in this context. Binary classification with unequal class importance is called \textit{asymmetric classification}, to address which two frameworks have been developed: \textit{cost-sensitive learning} \cite{elkan2001foundations} and \textit{Neyman-Pearson classification}\footnote{The Neyman-Pearson classification inherits its name from the Neyman-Pearson lemma due to a similar asymmetric nature: minimizing one type of error while controlling the other type of error.} \cite{cannon2002learning, scott2005neyman,tong2018neyman}. Specifically, the cost-sensitive learning framework achieves a small false negative rate by placing on it a large weight relative to the weight on the false positive rate in the objective function; the Neyman-Pearson classification framework guarantees a high-probability control on the population-level false negative rate while minimizing the false positive rate. Another two commonly-used accuracy measures for binary classification are the \textit{area under a receiver operating characteristic curve} (AUROC) and the \textit{area under a precision-recall curve} (AUPRC)\footnote{Important properties and dinstinctions between AUROC and AUPRC include but are not limited to: AUROC is invariant to the population sizes of the two classes, while AUPRC is not; AUROC can be overly optimistic if the two classes are extremely imbalanced in training data, while AUPRC does not have this issue. For detailed information, please refer to \cite{he2013imbalanced, branco2016survey, fernandez2018learning}.}. However, these two measures are not evaluation criteria for one decision rule (classifier) but rather evaluate a trained classification algorithm (e.g., logistic regression with parameters estimated from training data) with varying decision thresholds, each of which corresponds to a decision rule.

In summary, the evaluation of decision rules in hypothesis testing is less straightforward than in binary classification. To choose a statistical test for a specific data set, users have to use subjective judgment to decide whether test assumptions are reasonably justified. On the other hand, classification algorithms can be compared on a more objective ground, the \textit{Common Task Framework} \cite{donoho201750}, of which influential examples include the Kaggle competitions \cite{taieb2014gradient, graham2015kaggle, iglovikov2017satellite, zou2017youtube, sutton2019crowd} and the DREAM challenges \cite{bansal2014community, eduati2015prediction, sieberts2016crowdsourced, guinney2017prediction, seyednasrollah2017dream, salcedo2020community}. The Common Task Framework consists of three essential elements: training data, competing prediction algorithms, and validation data. A comparison is considered fair if all competing algorithms use the same training data to construct decision rules, which are subsequently evaluated on the same validation data using the same prediction accuracy measure.

\paragraph{Sample sizes vs. evaluation criteria.} A general principle in data sciences is, if a sample is unbiasedly drawn from a population, the larger the sample size, the more information we have about the population. This \textit{large-sample principle} holds for both hypothesis testing and binary classification; for example, data from a larger number of representative patients would lead to better decision rules for both biomarker detection and disease diagnosis. However, between the two strategies there is an interesting but often neglected distinction: from a population with finite instances (e.g., the human population), the largest possible sample, which is equivalent to the whole population, would make a valid statistical test achieve a perfect power given any significance level $\alpha$, while the largest possible training data set might not lead to a classification rule with perfect prediction accuracy. While this distinction is fundamentally rooted in mathematics, an intuitive understanding can be obtained from our biomarker detection and disease diagnosis examples. Imagine that we have measured everyone in the world. If a gene is indeed a disease biomarker, we can for sure see a difference in this gene's expression between all the people carrying this disease and the rest of the population, achieving the perfect power. On the other hand, diseased patients and undiseased individuals may not be perfectly separated by diagnostic measurements. That is, two patients may have similar symptoms and lab test results, but one patient is diseased and the other is not. When this happens, even if we have training data from all but one person in the world, we still cannot be 100\% sure whether the left-out individual has the disease just based on his or her diagnostic measurements. 

Table~\ref{tab:distinctions} summarizes the above distinctions between hypothesis testing and binary classification.

\begin{table}[htbp]
 \caption{Side-by-side comparison of hypothesis testing and binary classification}
  \centering
  \begin{tabular}{p{5.5cm}lp{5cm}}
    \toprule
         & Hypothesis testing     & Binary classification \\
    \toprule
   Symmetry between binary answers    & Asymmetric (default is 0) & Symmetric or asymmetric      \\
    \midrule
    \# of instances to make one decision given a decision rule    & \multirow{2}{*}{$\ge 1$ (the larger the better)} & \multirow{2}{*}{$ = 1$}      \\
    \midrule
    Available binary answers & No  & Yes (training data)    \\
    \midrule
    Evaluation criteria     & Power (given a significance level)       & Prediction accuracy  \\
    \midrule
    \multirow{2}{*}{With the largest possible \# of instances}     & \multirow{2}{*}{Power $ = 1$}       & Prediction accuracy not necessarily perfect  \\
    \bottomrule
  \end{tabular}
  \label{tab:distinctions}
\end{table}

\section{A checklist of five practical guidelines for choosing between hypothesis testing and binary classification}
\label{sec:guidelines}

\begin{figure}[htbp]
    \centering
    \includegraphics[width=\textwidth]{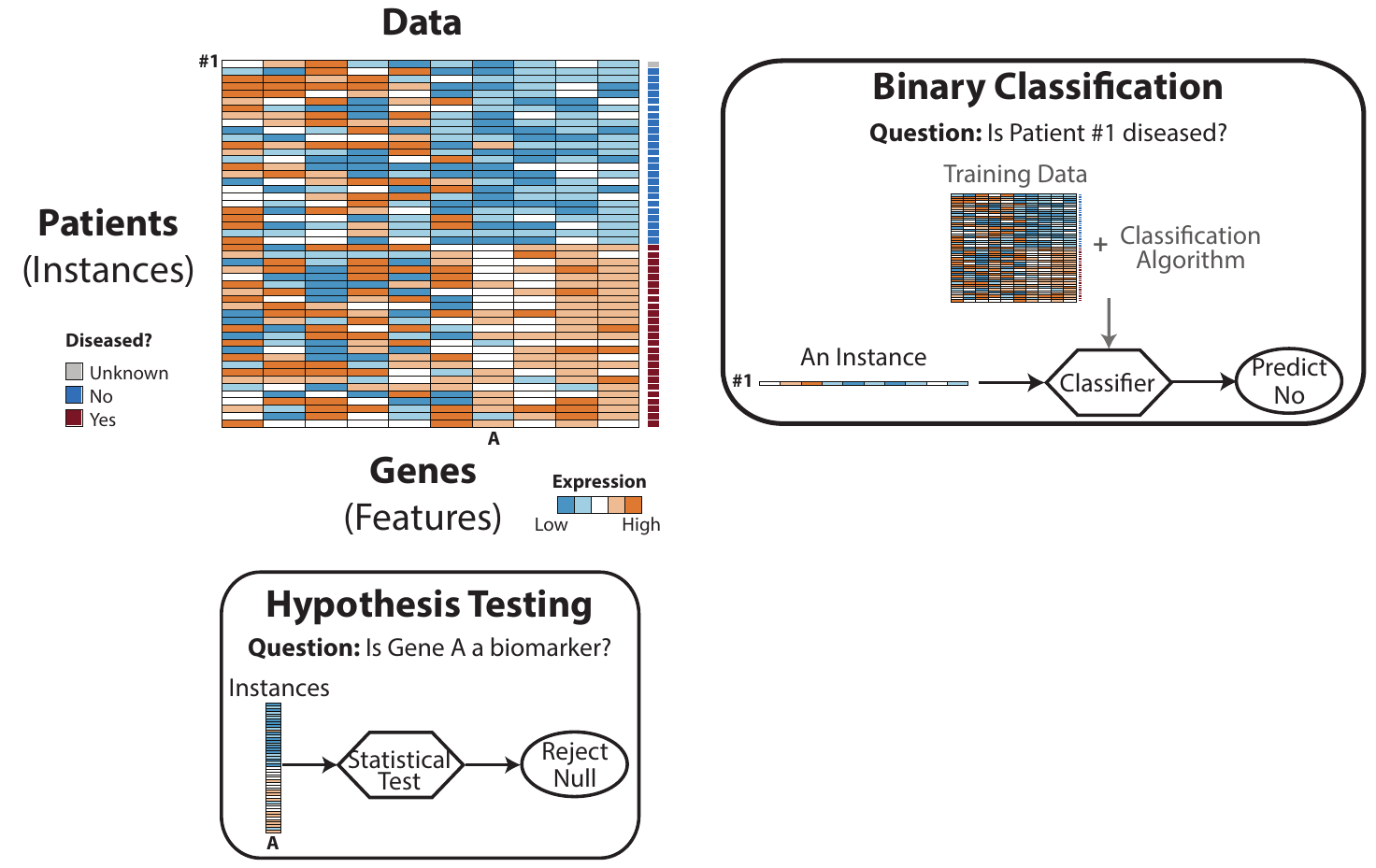}
    \caption{Illustration of a gene expression data set and two questions to be addressed by hypothesis testing and binary classification, respectively.}
    \label{fig:guidelines}
\end{figure}

Based on the key distinctions between hypothesis testing and binary classification, we provide a checklist of five practical guidelines for data analysts to choose between the two strategies. Figure~\ref{fig:guidelines} provides an illustration.

\paragraph{Guideline 1: Decide on instances and features.} Given a tabular data set, the first and necessary step is to decide whether rows and columns should be considered as instances and features respectively, or vice versa. The decision may seem trivial to experienced data scientists when columns represent variables in different units, e.g., gender, age, and body mass index, in which case columns should be considered as features for sure. However, the decision may become not-so-obvious in certain cases. For example in Figure~\ref{fig:guidelines}, a gene expression data set has rows and columns corresponding to patients and genes respectively, and all data values are in the same unit. The question is: should we consider patients as instances or features? To answer this question, the key is to understand instances as either (1) repeated measurements in the data collection process, or (2) a random sample from a population. Gene expression data are collected to understand gene expression patterns in healthy and diseased human sub-populations, so healthy and diseased patients are considered two random samples, each from one sub-population and satisfy (2). Hence, we conclude that patients are instances and genes are features. In general, the answer depends on the experimental design and the scientific question, both of which will determine what the underlying population is, as we will see in the cancer driver gene prediction example (Section~\ref{sec:example}).

\paragraph{Guideline 2: List the binary decisions to be made.} The second step is to outline the binary decisions to be made from the data. Formulate analytical tasks such as biomarker detection and disease diagnosis into binary questions, for which binary decisions will be made.  Divide binary questions into those related to features and others concerning instances. For example, whether a gene is a biomarker is a feature-related question, and whether a patient has a disease is an instance-related question. Hypothesis testing can only answer feature-related questions, while binary classification can only address instance-related questions. 

\paragraph{Guideline 3: Assess the availability of known binary answers in data.} After a list of binary questions is at hand, the next question is: do the data contain any known answers? If we already have an answer to a binary question, we cannot formulate that question as a hypothesis testing task. In the case where some instances contain known binary labels but we concern about the unknown labels of the rest of instances, we are facing a binary classification task, just as in disease diagnosis. Otherwise, if the data contain no binary labels, we do not have training data to construct a classifier, which, if not given, would prohibit us from predicting unknown labels of instances. 

\paragraph{Guideline 4: Count the number of instances for making each binary decision.} Suppose that we are given a decision rule, i.e., a statistical test or a classifier in the form of a formula or a computer program that can take our data as input and output a binary answer. An easy check is to count the number of input instances needed to output each binary decision. If we are expecting one decision per input instance, it is likely a binary classification task. Otherwise, if each binary decision needs to be made from a group of instances together, the task cannot be binary classification but might be formulated as hypothesis testing.

\paragraph{Guideline 5: Evaluate the nature of binary questions.} The most fundamental check is to evaluate each binary question by its nature: is the question regarding the unseen population of which our observed instances are a subset or regarding a particular instance? Asking whether a gene is a disease biomarker is a question of the former type, as it concerns whether this gene distinguishes the human sub-population with the disease from the rest of the population. In contrast, asking whether an individual has the disease is a question of the latter type. Hypothesis testing and binary classification are designed for answering questions of the former and latter types, respectively, as shown in Figure~\ref{fig:guidelines}. 

We recommend practitioners to check all the five guidelines before deciding whether hypothesis testing or binary classification is the correct strategy for a data analysis question.

\section{Cancer driver gene prediction: hypothesis testing or binary classification?}\label{sec:example}
Lastly, we present an important application example of cancer driver gene prediction to illustrate the distinction between hypothesis testing and binary classification. We will try to avoid using technical terms as much as possible for the ease of general readers. In the problem of cancer driver gene prediction, the goal is to utilize an individual gene's mutational signatures\footnote{Mutational signatures are summarized from multiple patient databases, and we do not have access to individual patient's data in this example.} such as the number of missense mutations to predict how likely the gene drives cancer. We have knowledge of a small set of cancer driver genes and neutral genes that are unlikely to drive cancer. The question is, how can we leverage this knowledge to predict whether a less-studied gene is a cancer driver gene? A famous algorithm, TUSON, addresses this question using a hypothesis testing approach \cite{davoli2013cumulative, tokheim2016evaluating}. Briefly, it regards mutational signatures as features and uses hypothesis testing to assess how much an individual gene resembles known neutral genes based on each feature: the gene's feature value is used as the test statistic, whose distribution under the null hypothesis (i.e., the gene is a neutral gene) is estimated from the feature values of known neutral genes; from the test statistic and the approximate null distribution, the gene receives a $p$-value for that feature. Suppose that there are ten features in total; then each gene receives ten $p$-values, which are subsequently combined into a single $p$-value by \textit{Fisher's method} \cite{fisher1992statistical}. 

From a statistical perspective, there are four apparent issues with this hypothesis testing approach. First, each hypothesis test, one per gene per feature, only utilizes the known neutral genes (to estimate the null distribution) but does not fully capture the valuable information in known cancer driver genes\footnote{To be exact, known cancer driver genes are used to select the predictive features before hypothesis testing is performed. However, in each test for one gene and one feature, the information of known cancer driver genes is only partially reflected in the direction of the gene's $p$-value: two-sided, larger-than, or smaller-than \cite{davoli2013cumulative}, which excludes the possibility that the known neutral genes may have feature values on the two sides of those of the known cancer driver genes.}. Second, each hypothesis test is performed using a sample of size one (i.e., the test statistic is the feature value of one gene), which is known to be not powerful and thus undesirable (i.e., if the gene is a cancer driver gene, we may miss it with a high chance). This is the reason why we recommend using more than one instance for hypothesis testing (Table~\ref{tab:distinctions} and Guideline 3). Third, combining multiple $p$-values into a single $p$-value is a difficult task, especially when $p$-values are not independent of each other. The fact that mutational signatures are observed to be correlated features, their resulting $p$-values are correlated for each gene, violating the independence assumption of Fisher's method. Although there are methods for combining dependent $p$-values \cite{brown1975400, kost2002combining, wilson2019harmonic, liu2020cauchy}, they cannot address the most fundamental question---what is the population behind each hypothesis test?---leading to the last issue. Fourth, the population behind the null hypothesis is unspecified: for a given test, is the population about that particular gene or all the genes? Therefore, we think that this hypothesis testing approach is inappropriate for this cancer driver gene problem, which, instead, should be formulated as a binary classification task for the following reasons.

Here we revisit this problem by following our checklist. Under Guideline 1, we consider genes as instances and mutational signatures as features, consistent with the existing studies. The reason is that we treat known cancer driver genes and neutral genes as a sample from the whole gene population of our interest, while we consider mutational signatures as given and we are not interested in the population they come from. Note that here genes are no longer treated as features as in biomarker detection where patients are instances. The contrast of the two examples suggests that a real quantity may be formulated as an instance or a feature depending on the data and the question of interest, and Guideline 1 provides a practical solution. Under Guideline 2, we conclude that the binary decisions to be made are instance-related because we would like to predict whether each gene is a cancer driver gene or not. Guideline 3 leads us to identify training data: known cancer driver genes and neutral genes. Next, if we already have a decision rule, we just need to input one gene to obtain its binary label: cancer driver gene or not. Hence, we only need one instance for each binary decision, concluding Guideline 4. Finally, we evaluate the nature of each binary question, as suggested by Guideline 5, and we can see that each question is only concerning one instance (gene), not the gene population. After checking all the five guidelines, it becomes evident that this cancer driver gene prediction problem is better suited to be addressed by binary classification.

Why does TUSON adopt the hypothesis testing approach? Its analysis results show that it aims to control the proportion of false discoveries among the predicted cancer driver genes, a criterion closely related to the FDR\footnote{The difference is that the FDR is the expected proportion of false discoveries among discoveries, where the expectation is taken over possible input data sets. However, this difference has been largely neglected in biomedical studies.}, which is widely used in multiple testing as we have discussed. Our guess is that the TUSON authors formulate cancer driver gene prediction as a multiple testing problem because they want to apply the famous Benjamini-Hochberg procedure to control the FDR by setting a cutoff on $p$-values, one per gene. However, this approach requires the validity of each $p$-value, which must follow a uniform distribution between zero and one under the null hypothesis. Due to the third issue we mentioned above (the violation of the assumption of Fisher's method), the combined $p$-value of each gene has no guarantee to satisfy this requirement. Here we would like to point out that the FDR concept is not restricted to multiple testing; in fact, it is a general evaluation criterion for multiple binary decisions, where each decision rule could be established by hypothesis testing or binary classification. Therefore, the goal of FDR control should not drive the choice between hypothesis testing and binary classification; instead, the choice should be based on the distinctions between the two strategies, as we have discussed in this work. Admittedly, the FDR has been rarely used as an evaluation criterion in binary classification; however, its closely-related criterion \textit{precision}\footnote{Precision is a criterion evaluated on a given set of validation data. It is a proportion, but unlike the FDR, it is not an expected proportion. How to implement a theoretically-guaranteed FDR control in binary classification is an open question for data science researchers.}, the proportion of correct predictions among all positive predictions, is widely used, such as in AUPRC. For cancer driver gene prediction, if we adopt the binary classification approach, we may compare competing classification algorithms by evaluating their AUPRC values using cross-validation. After we choose the algorithm that achieves the largest AUPRC value, we can train it on known cancer driver genes and neutral genes using their mutational signatures, and we can set a threshold on the trained algorithm based on our desired precision level to obtain a classifier (decision rule). Then we can simply apply the classifier to predict whether a less-studied gene is a cancer driver gene from its mutational signatures. In fact, we have implemented this approach and shown that it leads to more accurate discoveries than previous studies do \cite{LYU2020.07.21.213702}.

\section{Discussion}

In summary, hypothesis testing and binary classification have been regarded as two separate topics that have rarely been compared with each other in data science education and research. However, their distinctions in applications are not as apparent as in methodological research, where instances and features are well defined from the beginning. Instead, in applications how to formulate real quantities into instances or features is always a challenging task, a reason that obscures the distinctions between the two strategies. In this work, we attempt to summarize and compare the two strategies for the broad scientific community and the data science industry, and we provide five practical guidelines to help data analysts better distinguish between the two strategies in data analysis.

As an extension, when instances are categorized into more than two classes or groups, similar distinctions exist between hypothesis testing and classification. Multi-group comparison is a typical hypothesis testing question. For example, if the question is whether a feature has the same expected value in all groups, it can be addressed by the analysis of variance (ANOVA). In contrast, multi-class classification seeks to assign every unlabeled instance one or more of the multiple class labels. In fact, compared to our previous discussion, here the distinction between hypothesis testing and classification is clearer: hypothesis testing still gives rise to a binary decision (reject or not the null hypothesis that a feature has the same expected value in all groups), while classification leads to a decision with more than two possible answers (every instance has more than two possible labels).

To conclude, we would like to emphasize again that hypothesis testing and binary classification are different in nature: the former concerns an unobservable population-level property of a feature, while latter pertains to an observable label of an instance. Despite this inherent difference, it is possible to adopt ideas from one strategy to develop new decision rules for the other strategy, or use one strategy as a preceding step to enable the application of the other strategy. In one direction, powerful test statistics with theoretical foundations in hypothesis testing may inform the construction of classifiers in binary classification. For example, the likelihood ratio test statistic bears a similar mathematical form as the na\"ive Bayes classifier\footnote{The key distinction is that the likelihood ratio test statistic takes all the available instances as input to construct a decision rule for one test, while the na\"ive Bayes classifier, if trained, takes one instance as input and outputs a predicted binary label.}. There are other concrete examples that leverage test statistics to construct classifiers \cite{Liao.Akritas.2007, He.Sheng.Liu.Zou.2019}. In the other direction, successful classification algorithms may motivate new scientific questions that can be investigated by hypothesis testing. For example, convolutional and recurrent neural networks have demonstrated superb capacity to extract predictive features from unstructured data such as images and texts \cite{krizhevsky2012imagenet, hu2014convolutional, mikolov2011extensions, sak2014long}. For those extracted features that are interpretable (such as a feature related to one brain region in fMRI images), researchers may want to investigate whether such a feature differs between two groups of images from different patient cohorts, and then they can employ hypothesis testing on that particular feature. Furthermore, in some special and rare examples, an algorithm may serve the purposes of both strategies. The most famous example is logistic regression, which is both a classification algorithm and a testing approach for deciding whether associations exist between features and binary labels. In a binary classification task whose goal is to label instances, logistic regression is used to construct a classifier. Meanwhile, logistic regression and its accompanying Wald test can also be used to investigate how each feature influences binary labels of instances \cite{mccullagh2018generalized}. Ultimately, effective data analysis requires appropriate usage of hypothesis testing and binary classification for suitable tasks, and this can only be realized when data analysts are well informed of the distinctions and connections between the two strategies.

\section{Acknowledgement}
We have received many insightful comments during the preparation of this manuscript. We thank Dr. Wei Li at University of California, Irvine for bringing the cancer-driver gene prediction problem to our attention and commenting on our manuscript. We also thank the following people for their comments: Heather J. Zhou, Tianyi Sun, Kexin Li, Wenbin Guo, Yiling chen, Dongyuan Song, and Ruochen Jiang in the Junction of Statistics and Biology (http://jsb.ucla.edu) at University of California, Los Angeles, Lijia Wang, Man Luo, and Dr. Michael S. Waterman at University of Southern California, Ruhan Dong at Pandora Media, Dr. Han Chen at University of Texas Health, Dr. Mark Biggin at Lawrence Berkeley National Laboratory, and Dr. Thomas Burger at the Université Grenoble Alpes and French National Centre for Scientific Research (CNRS).

\bibliographystyle{unsrt}  
\bibliography{references}

\end{document}